\documentclass[a4paper,11pt]{article}
\usepackage{cite}
\usepackage{jheppub} 
\usepackage{etoolbox}
\patchcmd{\maketitle}{\@fpheader}{}{}{}
\usepackage[T1]{fontenc} 
\usepackage[utf8]{inputenc} 
\usepackage{microtype} 
\usepackage{lmodern} 
\usepackage{mdframed} 
\usepackage{mathtools}
\usepackage{graphicx}
\usepackage{cancel} 
\usepackage{todonotes} 
\usepackage[normalem]{ulem} % either use this (simple) or
\usepackage{soul} % use this (many fancier options)
\usepackage{amsmath}
\usepackage{amsfonts}
\usepackage{amssymb}
\usepackage{stmaryrd}
\usepackage[mathscr]{euscript}
\usepackage{mathtools}
\usepackage{mathrsfs}
\usepackage{multicol}
\usepackage{soul}
\usepackage{mleftright} \mleftright
\usepackage{tensor} 
\usepackage{braket} 
\usepackage{mathrsfs} 
\usepackage{bbold}
\usepackage{color}
\usepackage{braket}
\usepackage{slashed}
\usepackage{hyperref}

\newcommand{\comment}[1]{}

\providecommand{\sc}{{ }}
\renewcommand{\sc}{{ }}
\DeclareMathAlphabet{\mathfs}{U}{rsfs}{m}{n}                     %

\newcommand{\be}{\nopagebreak[3]\begin{equation}}
\newcommand{\ee}{\end{equation}}
\newcommand{\bee}{\nopagebreak[3]\begin{equation*}}
\newcommand{\eee}{\end{equation*}}
\newcommand{\ba}{\nopagebreak[3]\begin{eqnarray}}
\newcommand{\ea}{\end{eqnarray}}
\newcommand{\baa}{\nopagebreak[3]\begin{eqnarray*}}
\newcommand{\eaa}{\end{eqnarray*}}
\newcommand{\bal}{\nopagebreak[3]\begin{aligned}}
\newcommand{\eal}{\end{aligned}}

\newcommand{\bseq}{\nopagebreak[3]\begin{subequations}}
\newcommand{\eseq}{\end{subequations}\noindent}

\title{AdS$_3$ Carroll gravity: asymptotic symmetries and C-thermal configurations}
\author[a,b]{Luis Avilés,}
\author[a,b,c]{Oscar Fuentealba,}
\author[d]{Diego Hidalgo}
\author[e,f]{and Pablo Rodríguez}

\affiliation[a]{Instituto de Ciencias Exactas y Naturales (ICEN), Universidad Arturo Prat,\\ Playa Brava 3256, 1111346 Iquique, Chile}
\affiliation[b]{Facultad de Ciencias, Universidad Arturo Prat,\\ Avenida Arturo Prat Chac\'on 2120, 1110939 Iquique, Chile}
\affiliation[c]{International Solvay Institutes, \\ ULB-Campus Plaine CP231, B-1050 Brussels, Belgium}
\affiliation[d]{Science Institute, University of Iceland,\\ Dunhaga 3, 107 Reykjav\'ik, Iceland}
\affiliation[e]{Departamento de Ciencias Exactas, Universidad de Los Lagos,\\ Av. Fuchslocher 1305, Osorno, Chile}
\affiliation[f]{Departamento de Ciencias Matemáticas y Físicas, Universidad Católica de Temuco,\\
Montt 56, Casilla 15-D, Temuco, Chile}

\emailAdd{luaviles@unap.cl}
\emailAdd{ofuentealba@unap.cl}
\emailAdd{dhidalgo@hi.is}
\emailAdd{pablo.rodriguez@ulagos.cl}

\preprint{{\bf } }

\abstract{The asymptotic structure of three-dimensional Carroll gravity with negative cosmological constant is studied. We formulate a consistent set of boundary conditions preserved by an infinite-dimensional extension of the AdS$_3$ Carroll algebra, which turns out to be isomorphic to a precise generalized BMS$_3$ algebra. This is described by four independent functions of the circle at infinity, generating spatial superrotations, Carroll superboosts, spatial supertranslations and time supertranslations. Remarkably, this asymptotic  symmetry algebra contains as subalgebras to BMS$_3$ (generated by spatial superrotations and time supertranslations) and the two-dimensional conformal algebra (spanned by spatial superrotations and spatial supertranslations). We also introduce a new solution -- endowed with a Carroll extremal surface -- that fulfills this set of asymptotic conditions. By taking advantage of the Chern-Simons formulation of the theory, Carroll thermal properties, obtained from regularity conditions, and entropy of the configuration are also addressed.}

\begin{document}
\maketitle
\flushbottom
\section{Introduction}

Carrollian symmetries have been shown to be at the core of recent and highly intriguing developments in theoretical physics. The Carroll group, discovered by  Lévy-Leblond and Sen Gupta \cite{Levy-Leblond:1965,SenGupta:1966qer} in the sixties, was first found by taking the limit where the speed of light tends to zero in the Poincar\'e group. Carroll group then took a new boost thanks to the observation by Duval, Gibbons, and Horvathy that the BMS$_{d+1}$ {algebra} is isomorphic to the conformal Carroll algebra in $d$ dimensions \cite{Duval:2014uva} (see also \cite{Bagchi:2010zz} where this isomorphism was previously observed in the particular case of BMS$_3$ and the conformal Galilean algebra in two dimensions\footnote{Carroll and {Galilei} algebras are isomorphic in two dimensions.}). Carrollian structures have shown to emerge in generic null hypersurfaces, as it is the case of the event horizon of black holes  \cite{Penna:2015gza,Penna:2018gfx,Donnay:2019jiz,Ciambelli:2019lap,Freidel:2022vjq,Bagchi:2023cfp,Bagchi:2024rje}, describing the asymptotic symmetries near a spacelike singularity \cite{Belinsky:1970ew,Belinsky:1982pk,Henneaux:1982qpq,Damour:2002et,Belinski:2017}, and more recently playing a key role in the context of four-dimensional flat space holography \cite{Ciambelli:2018wre,Donnay:2022aba,Bagchi:2022emh,Campoleoni:2022wmf,Donnay:2022wvx,Bagchi:2023fbj,Saha:2023hsl,Salzer:2023jqv,Nguyen:2023miw,Bagchi:2023cen,Mason:2023mti,Bekaert:2024itn,Bagchi:2024efs}.

One may wonder about Carroll invariant field theories {and whether} this could be obtained by limiting processes from Lorentz invariant theories. Hamiltonian \cite{Henneaux:2021yzg} and Lagrangian \cite{deBoer:2021jej,Hansen:2021fxi,Bergshoeff:2022qkx} approaches allowed to foresee two inequivalent possible limits, namely, one electric and one magnetic. Notably, as discussed in \cite{Henneaux:2021yzg}, all Carroll field theories (independently of their type) share the common feature of possessing a vanishing Poisson bracket between energy densities, which was first noted in the context of zero signature gravity in \cite{Teitelboim:1978wv,Henneaux:1979vn} (see also \cite{Isham:1975ur}). Carroll field theories have also been constructed through an intrinsic viewpoint (no limiting process), as, for instance, the case of a scalar field with non-trivial dynamics \cite{Ciambelli:2023xqk}, Carroll swiftons \cite{Ecker:2024lur} (fields with propagation outside the Carroll lightcone) and by (finite) deformation processes from conformal field theories \cite{Rodriguez:2021tcz,Tempo:2022ndz,Parekh:2023xms,Bagchi:2022nvj,Bagchi:2024unl}.

A profound understanding of a theory requires deepening into its asymptotic structure. In this sense, the study of the asymptotic symmetries of Carrollian theories of gravity in four dimensions have also been addressed in the literature, specifically, in both electric and magnetic Carrollian limits of Einstein gravity in  \cite{Perez:2021abf,Perez:2022jpr} and its coupling to a Yang-Mills field \cite{Fuentealba:2022gdx}. The question of studying their three-dimensional analogues is then direct. Particularly, our motivation relies on the fact that degeneracy of the structures describing Carroll geometries leads to infinite-dimensional symmetries (see e.g., \cite{Blau:2015nee}). Then, exploring the richness of the asymptotic structure and the searching for interesting Carrollian configurations in the context of a controlled set-up (as AdS$_3$ Carroll gravity) are worthy endeavours.

We will make use of the Chern-Simons formulation of AdS$_3$ Carroll gravity \cite{Matulich:2019cdo,Ravera:2019ize} (see \cite{Bergshoeff:2016soe} for the flat case) and thus explore the existence of novel infinite-dimensional extensions of the AdS$_3$ Carroll algebra. This question has been previously analysed in the context of a vanishing cosmological constant in \cite{Bergshoeff:2016soe}, where asymptotic symmetries were generated by spatial rotations, time translations, and infinite-dimensional extensions of both Carroll boosts and spatial translations. The isomorphism between AdS-Carroll and Poincar\'e algebras \cite{Bacry:1968zf,Bergshoeff:2022eog} was exploited very recently in \cite{Aviles:2024llx} with the aim of {formulating} a consistent set of boundary conditions that is preserved under BMS$_3$, in the presence of a negative cosmological constant.

In this paper, we introduce an inequivalent set of boundary conditions, which turns out to be preserved by an infinite-dimensional extension of the AdS$_3$ Carroll algebra, described by \textit{four} independent functions of the circle at infinity, to wit, spatial superrotations, Carroll superboosts, time supertranslations and spatial supertranslations. This algebra is isomorphic to a precise generalized BMS$_3$ algebra of \cite{Caroca:2017onr}, obtained through the semi-group expansion method (originally developed in \cite{Izaurieta:2006zz}). The formulation of the theory in terms of gauge fields allowed us to study the thermodynamics of the configurations along the lines of \cite{Bunster:2014mua} by exploiting the techniques developed in the context of higher spin extensions of three-dimensional gravity (see, e.g. \cite{Perez:2014pya,Bunster:2014mua} and references therein), and the concepts introduced in \cite{Ecker:2023uwm}, defining Carroll thermal (C-thermal) manifolds. The C-thermal properties of the found solution resemble the ones of the asymptotically flat cosmological configuration of three-dimensional gravity \cite{Ezawa:1992nk,Cornalba:2002fi,Cornalba:2003kd,Barnich:2012xq} (see also \cite{Gary:2014ppa,Matulich:2014hea}). From this perspective, this C-thermal manifold could then be regarded as Carrollian analogue of a cosmological configuration.

The plan of the paper goes as follows: In Section \ref{sec:action}, we recall how AdS$_3$ Carroll gravity can be recast as a Chern-Simons theory for the AdS Carroll group in three dimensions. The connection of the gauge field formulation with Carroll structures is also shown. In Section \ref{sec:asymptotic-structure}, we provide a consistent set of boundary conditions preserved by an infinite-dimensional extension of the AdS$_3$ Carroll algebra. The flat limit  ($\ell \rightarrow \infty$, where $\ell$ is the AdS radius) is directly taken in every step. This allows us to obtain an infinite-dimensional extension of the flat Carroll algebra in three dimensions. We briefly address the emergence of the Heisenberg current algebra and its connection with the so-called BMS-like algebras in the context of Carroll gravity. In Section \ref{sec:thermo}, we describe a new solution of the theory included within this set of asymptotic conditions. Specifically, we determine the Carroll structures associated {with} a configuration which turns out to be endowed with a Carroll extremal surface. By taking advantage of the Chern-Simons formulation, we perform an analysis of the thermodynamics of this configuration in terms of gauge fields. This is {done} by requiring a trivial holonomy around a thermal cycle, which fixes the value of the chemical potentials in terms of the global charges and then computing the entropy. Finally, Section \ref{sec:endings} is devoted to the concluding remarks.

\section{AdS$_3$ Carroll gravity as a Chern-Simons theory}\label{sec:action}

Carroll gravity in three spacetime dimensions in {the} presence of a negative cosmological constant $\Lambda=-1/\ell^2$ can be recast as a Chern-Simons theory for the AdS$_3$ Carroll {algebra}   \cite{Matulich:2019cdo,Ravera:2019ize,Bergshoeff:2016soe,Aviles:2024llx}. The nonvanishing commutators of the AdS$_3$ Carroll algebra read
\begin{align}\label{eq:AdSCarrollalgebra}
\big[J,P_{a}\big]	&=\epsilon_{ab}P^{b}\,,\hspace{1.1cm} \big[J,C_{a}\big]=\epsilon_{ab}C^{b}\,,\quad\big[C_{a},P_{b}\big]=\delta_{ab}H\,,\\
\big[P_{a},P_{b}\big]	&=-\frac{1}{\ell^{2}}\epsilon_{ab}J\,,\quad\big[P_{a},H\big]=-\frac{1}{\ell^{2}}C_{a}\,.
\end{align}
The set $(J, C_a)$ generates the homogeneous Carroll group, $J$ stands for the spatial rotation generator, and $C_a$ (with $a=1,2$) corresponds to the Carroll boost generators. Time and spatial translations are generated by $H$ and $P_a$, respectively. The orientation is chosen in such a way that $\epsilon_{12}=1$. {The indices are raised and lowered by the two-dimensional Kronecker delta $\delta_{ab}$}. For simplicity, we choose the invariant bilinear product with nonvanishing components
\begin{equation}\label{eq:bracket}
\langle J\,,H\rangle=1\,,\quad  \langle C_{a}\,,P_{b}\rangle=-\epsilon_{ab}\,.
\end{equation}
The field content can be organized in the Lie-algebra valued one-form
\begin{equation}\label{eq:gauge-connection}
A=\tau H+e^aP_a+\omega J+\omega^a C_a\,,
\end{equation}
where $\tau$ and $e^a$ stand for the local vielbeins, while $\omega$ and $\omega^a$ correspond to the Carroll spin connections.

Given the previous definitions, we can now write the action for the theory as the Chern-Simons form 
\begin{equation}
I=\frac{k}{4\pi}\int \Bigg\langle AdA+\frac{2}{3}A^3 \Bigg\rangle\,,
\end{equation}
which, {by using \eqref{eq:AdSCarrollalgebra} and \eqref{eq:bracket}, can be written explicitly as}
\begin{equation}\label{eq:action}
I=\frac{k}{2\pi}\int\left(\epsilon_{ab}e^aR^b-\tau R+\frac{1}{\ell^2}\epsilon_{ab}\tau e^ae^b\right)\,,
\end{equation}
where the Chern-Simons level reads $k=1/4G_C$, with $G_C=c^{-1}G$ \cite{Aviles:2024llx} (where $c$ is the speed of light and $G$ stands for the Newton gravitational constant). Equations of the motion ask for the vanishing of the two-form field strength
\begin{equation}
F=dA+A^2=T^aP_a+T H+R J+R^a C_a\,,
\end{equation}
where
\begin{align}
T_a&=de_a+\epsilon_{ab}e_b \omega\,,\\
T&=d\tau+\omega_ae^a\,,\\
R_a&=d\omega_a+\frac{1}{\ell^2}\tau e_a-\epsilon_{ab}\omega \omega_b\,,\\
R&=d\omega-\frac{1}{2\ell^2}\epsilon_{ab}e^ae^b\,.
\end{align}

{For forthcoming purposes, we introduce some basics about Carroll geometry.} Carroll structures are given by the degenerate Carroll metric
\begin{equation}
g_{\mu\nu}=\delta_{ab}\,e^a_{\mu} \, e^b_{\nu}\,,
\end{equation}
and a vector $n^{\mu}$, satisfying the orthogonality condition  
\begin{equation}
g_{\mu\nu}n^{\mu}=0\,.
\end{equation}
The components of the orthogonal vector $n^\mu$ can be obtained by solving the following completeness conditions that determine the dual basis of the tangent space formed by $E^{\mu}_a$ and $n^{\mu}$ \cite{Bergshoeff:2014jla} (see also \cite{Bergshoeff:2017btm,Campoleoni:2022ebj}) 
\begin{equation}\label{eq:completeness}
e^a_{\mu}E^{\mu}_b=\delta^a_b\,,\quad \tau_{\mu}n^{\mu}=1\,,\quad n^{\mu}e^a_{\mu}=0\,,\quad e^a_{\mu}E^{\nu}_{a}+\tau_{\mu}n^{\nu}=\delta^{\nu}_{\mu}\,.
\end{equation}
From here, we can see that Carroll structures possess a freedom\footnote{Special thanks to Prof. Marc Henneaux for this point.}. Given the local Carroll invariant structures $g_{\mu\nu}$ and $n^{\mu}$, we cannot determine completely the tetrad  $E^{\mu}_a$ and the one-form $\tau_{\mu}$ (these could differ by the shifts $E^{\mu}_a\rightarrow E^{\mu}_a+f_a n^{\mu}$ and $\tau_{\mu}\rightarrow \tau_{\mu}+f_a e^{a}_{\mu}$, parametrized by an arbitrary function $f_a$). This freedom is frozen if instead we initially give the cotangent space basis $\tau_{\mu}$ and $e^{a}_{\mu}$, and then we determine the tangent space dual basis through the completeness conditions. In this case, we shall directly read the cotangent fields from the gauge connection \eqref{eq:gauge-connection}.

The action principle that is obtained in terms of Carroll structures from the Chern-Simons action \eqref{eq:action} turns out to be the magnetic limit of three-dimensional Einstein gravity \cite{Aviles:2024llx}. A gauging procedure for the three-dimensional Carroll algebra that connects a first-order formulation with the electric limit of Einstein gravity was proposed in \cite{Hartong:2015xda} (see also  \cite{Pekar:2024ukc} for the Cartan-like formulation of electric Carrollian gravity in four dimensions).

\section{Asymptotic structure}\label{sec:asymptotic-structure}
\subsection{Asymptotic conditions}
\label{sec:asymptotics}

In order to find a consistent set of boundary conditions, we follow the general lines given in \cite{Bunster:2014mua}, that is to say, $i)$ boundary conditions are chosen such that the symmetry group preserving them includes the AdS$_3$ Carroll group  (isometry subgroup of the vacuum), $ii)$ the fall-off is relaxed enough to capture interesting bosonic solutions, and $iii)$  the fall-off is also fast enough to ensure integrability and finiteness of the global charges. Taking into account these criteria, we propose the following boundary conditions
\begin{equation}
A(t,r,\phi)=b^{-1}a(t,\phi)b+b^{-1}db\,.
\end{equation} 
The radial dependence of the gauge connection is entirely captured by the permissible group element $b=e^{\alpha (r)P_2}$, where $\alpha(r)$ is a function of the radial coordinate to be written in section \ref{subsec:Configuration}. The auxiliary gauge field $a$ depends only on time $t$ and the angular coordinate $\phi$. Its nonvanishing components read
\begin{align}
a_{\phi}&=e^{\mathcal{S}}J+\mathcal{M}C_{1}+\mathcal{N}P_{1}+\mathcal{J}H\,, \label{eq:aphi}\\ 
a_t&=\Lambda[\mu_{\mathcal L},\mu_{\mathcal P},\mu_{\mathcal C},\mu_{\mathcal H}]\,,\label{eq:at}
\end{align}
where
\begin{eqnarray}
\nonumber \Lambda[\mu_{\mathcal L},\mu_{\mathcal P},\mu_{\mathcal C},\mu_{\mathcal H}]&=&e^{\mathcal{S}}\mu_{\mathcal{L}}J+e^{\mathcal{S}}\mu_{\mathcal{P}} P_2+\lambda_C C_2+\lambda_{H}H\\
&&-\left(e^{-\mathcal{S}}\lambda_C'-\mathcal{M}\mu_{\mathcal{L}}+\frac{1}{\ell^2}\mathcal{J}\mu_{\mathcal{P}}\right)C_1-\left(\mu_{\mathcal P}'+\mathcal{S}'\mu_{\mathcal P}-\mathcal N\mu_{\mathcal L}\right)P_1\,,
\end{eqnarray}
with
\begin{align}
\lambda_{C} & =\mu_{\mathcal C}'+\mathcal S'\mu_{\mathcal C}-\frac{1}{\ell^2}\mathcal N\mu_{\mathcal H}\,, \label{eq:lambdaC}\\
\lambda_{H} & =\mathcal J\mu_{\mathcal L}-\mathcal M\mu_{\mathcal P}-e^{-\mathcal S}\left(\mathcal N\mu_{\mathcal C}\right)'+e^{-\mathcal S}\left[\mu_{\mathcal H}''+\mathcal S'\mu_{\mathcal H}'+\left(e^{2\mathcal S}+\mathcal S''\right)\mu_{\mathcal H}\right]\,.\label{eq:lambdaH}
\end{align}
Here prime stands for derivative with respect to $\phi$. The dynamical fields $\mathcal{S}$, $\mathcal{M}$, $\mathcal{N}$ and $\mathcal{J}$ are arbitrary functions on $t$ and $\phi$. The functions $\mu_{\mathcal L}$, $\mu_{\mathcal P}$, $\mu_{\mathcal H}$ and $\mu_{\mathcal C}$ -- being also arbitrary functions on $t$ and $\phi$ -- introduced along the time component of the connection $a_t$, are also called chemical potentials, as they usually appear as conjugate of the global charges in the first law of thermodynamics.  The specific form of the time component of the connection is obtained from asking for the most general form consistent with the preservation of the boundary conditions under the asymptotic symmetry group \cite{Henneaux:2013dra}. The presence of the chemical potentials in the boundary conditions plays a key role when analyzing smoothness of the configurations \cite{Bunster:2014mua}. In this work, we will consider the chemical potentials fixed without variation. Other choices involving a functional dependence of the chemical potentials on the dynamical fields have been explored in the literature, which makes interesting connections with integrable systems, as discovered in \cite{Perez:2016vqo} for the case of the KdV hierarchy (see also \cite{Fuentealba:2017omf,Cardenas:2021vwo,Lara:2024cie}). 

We proceed to establish the preservation under the action of a Lie-algebra parameter $\lambda=\lambda(t,\phi)$ of the auxiliary gauge field,
\begin{equation}\label{eq:transflaw}
\delta_{\lambda} a=d\lambda+[a,\lambda]\,.
\end{equation} 
The  angular component $a_{\phi}$ is preserved provided the parameter takes the form
\begin{equation}
\lambda=\Lambda[\varepsilon_{\mathcal L},\varepsilon_{\mathcal P},\varepsilon_{\mathcal C},\varepsilon_{\mathcal H}]\,.
\end{equation}
Transformation laws of the dynamical fields (obtained from \eqref{eq:transflaw}) can be conveniently written as
\begin{align}
\delta \mathcal S&=\varepsilon_{\mathcal L}'+\mathcal S'\varepsilon_{\mathcal L}-\frac{1}{\ell^2}\mathcal{N}\varepsilon_{\mathcal P}\,,\\
\delta \mathcal{M}&=-\left(e^{-\mathcal{S}}\epsilon_C'-\mathcal{M}\varepsilon_{\mathcal L}+\frac{1}{\ell^2}\mathcal{J}\varepsilon_{\mathcal P}\right)'-e^{\mathcal{S}}\epsilon_C-\frac{1}{\ell^2}\left[\mathcal{N}\epsilon_H+\mathcal{J}\left(\varepsilon_{\mathcal P}'+\mathcal S'\varepsilon_{\mathcal P}-\mathcal{N}\varepsilon_{\mathcal L}\right)\right]\,,\\
\delta \mathcal{N}&=-\left(\varepsilon_{\mathcal P}'+\mathcal S'\varepsilon_{\mathcal P}-\mathcal{N}\varepsilon_{\mathcal L}\right)'-e^{2\mathcal{S}}\varepsilon_{\mathcal P}\,,\\
\delta \mathcal{J}&=\epsilon_H'-\mathcal{M}\left(\varepsilon_{\mathcal P}'+\mathcal S'\varepsilon_{\mathcal P}\right)+\mathcal{N}\left(e^{-\mathcal{S}}\epsilon_C'+\frac{1}{\ell^2}\mathcal{J}\varepsilon_{\mathcal P}\right)\,,
\end{align}
where
\begin{align}
\epsilon_{C} & =\varepsilon_{\mathcal C}'+\mathcal S'\varepsilon_{\mathcal C}-\frac{1}{\ell^2}\mathcal N\varepsilon_{\mathcal H}\,,\\
\epsilon_{H} & =\mathcal J\varepsilon_{\mathcal L}-\mathcal M\varepsilon_{\mathcal P}-e^{-\mathcal S}\left(\varepsilon_{\mathcal N\mathcal C}\right)'+e^{-\mathcal S}\left[\varepsilon_{\mathcal H}''+\mathcal S'\varepsilon_{\mathcal H}'+\left(e^{2\mathcal S}+\mathcal S''\right)\varepsilon_{\mathcal H}\right]\,.
\end{align}
The time component $a_t$ is preserved provided
\begin{align}\label{eq:eqmotion1}
\dot{ \mathcal S}&=\mu_{\mathcal L}'+\mathcal S'\mu_{\mathcal L}-\frac{1}{\ell^2}\mathcal{N}\mu_{\mathcal P}\,,\\
\dot{ \mathcal{M}}&=-\left(e^{-\mathcal{S}}\lambda_C'-\mathcal{M}\mu_{\mathcal L}+\frac{1}{\ell^2}\mathcal{J}\mu_{\mathcal P}\right)'-e^{\mathcal{S}}\lambda_C-\frac{1}{\ell^2}\left[\mathcal{N}\lambda_H+\mathcal{J}\left(\mu_{\mathcal P}'+\mathcal S'\mu_{\mathcal P}-\mathcal{N}\mu_{\mathcal L}\right)\right]\,,\\
\dot{  \mathcal{N}}&=-\left(\mu_{\mathcal P}'+\mathcal S'\mu_{\mathcal P}-\mathcal{N}\mu_{\mathcal L}\right)'-e^{2\mathcal{S}}\mu_{\mathcal P}\,,\\
\dot{  \mathcal{J}}&=\lambda_H'-\mathcal{M}\left(\mu_{\mathcal P}'+\mathcal S'\mu_{\mathcal P}\right)+\mathcal{N}\left(e^{-\mathcal{S}}\lambda_C'+\frac{1}{\ell^2}\mathcal{J}\mu_{\mathcal P}\right)\, \label{eq:eqmotion2},
\end{align}
together with the following suitable conditions on time derivatives of the gauge parameters
\begin{align}
\dot{\varepsilon}_{\mathcal L}&=\mu_{\mathcal L}\varepsilon_{\mathcal L}'-\mu_{\mathcal L}'\varepsilon_{\mathcal L}+\frac{1}{\ell^2}\left(\mu_{\mathcal P}\varepsilon_{\mathcal P}'-\mu_{\mathcal P}'\varepsilon_{\mathcal P}\right)\,,\\
\dot{\varepsilon}_{\mathcal P}&=\mu_{\mathcal P}\varepsilon_{\mathcal L}'-\mu_{\mathcal P}'\varepsilon_{\mathcal L}+\mu_{\mathcal L}\varepsilon_{\mathcal P}'-\mu_{\mathcal L}'\varepsilon_{\mathcal P}\,,\\
\dot{\varepsilon}_{\mathcal C}&=\mu_{\mathcal L}\varepsilon_{\mathcal C}'-\mu_{\mathcal L}'\varepsilon_{\mathcal C}+\mu_{\mathcal C}\varepsilon_{\mathcal L}'-\mu_{\mathcal C}'\varepsilon_{\mathcal L}+\frac{1}{\ell^2}\left(\mu_{\mathcal P}\varepsilon_{\mathcal H}'-\mu_{\mathcal P}'\varepsilon_{\mathcal H}+\mu_{\mathcal H}\varepsilon_{\mathcal P}'-\mu_{\mathcal H}'\varepsilon_{\mathcal P}\right)\,,\\
\dot{\varepsilon}_{\mathcal H}&=\mu_{\mathcal P}\varepsilon_{\mathcal C}'-\mu_{\mathcal P}'\varepsilon_{\mathcal C}+\mu_{\mathcal L}\varepsilon_{\mathcal H}'-\mu_{\mathcal L}'\varepsilon_{\mathcal H}+\mu_{\mathcal H}\varepsilon_{\mathcal L}'-\mu_{\mathcal H}'\varepsilon_{\mathcal L}+\mu_{\mathcal C}\varepsilon_{\mathcal P}'-\mu_{\mathcal C}'\varepsilon_{\mathcal P}\,.
\end{align}
The functions $\lambda_C$ and $\lambda_H$ are given in \eqref{eq:lambdaC} and \eqref{eq:lambdaH}, respectively.

Once we have ensured the preservation of our boundary conditions, we proceed to compute the global charges and their corresponding algebra.

\subsection{Infinite-dimensional extension of the AdS$_3$ Carroll algebra} \label{subsec:algebra}

The global charges for a Chern-Simons theory are determined from the surface integral of the canonical generator, which is obtained by following the Regge-Teitelboim approach \cite{Regge:1974zd}. The variation of the surface integral reads
\begin{equation}
\delta Q_{\lambda}=-\frac{k}{2\pi}\oint d\phi\, \left\langle\lambda, \delta a_{\phi}\right\rangle\,.
\end{equation}
Direct implementation of the boundary conditions given in the previous subsection yields the following surface integral
\begin{equation}\label{eq:charge}
\delta Q_\lambda =-\frac{k}{2\pi}\oint d\phi \, \left(\varepsilon_{\mathcal{L}}\delta\mathcal{L}-\varepsilon_{\mathcal{P}}\delta\mathcal{P}+\varepsilon_{\mathcal{C}}\delta\mathcal{C}-\varepsilon_{\mathcal{H}}\delta\mathcal{H}\right)\,,
\end{equation}
where 
\begin{eqnarray} 
\mathcal{L}	&=& e^{\mathcal{S}}\mathcal{J}\,,\label{eq:mathcalL} \\
\mathcal{P}	&=& e^{\mathcal{S}}\mathcal{M} \,, \\
\mathcal{C}	&=& \mathcal{N}\mathcal{S}'-\mathcal{N}'\,, \\
\mathcal{H}	&=& \frac{1}{2}\left[\frac{\mathcal{N}^{2}}{\ell^{2}}-e^{2\mathcal{S}}-2\mathcal{S}''+\left(\mathcal{S}'\right)^{2}\right]\label{eq:mathcalH}\,.
\end{eqnarray}
From \eqref{eq:charge}, we can see that the charge is readily integrable if we assume that the gauge parameters are field-independent. One can check that the charge in  \eqref{eq:charge} is conserved by virtue of the conditions on the time derivatives of the gauge parameters and the equations of the motion
\begin{align}
\dot{\mathcal{L}}&=2\mathcal L \mathcal \mu_{\mathcal L}'+\mathcal L'\mu_{\mathcal L}-2\mathcal P\mathcal \mu_{\mathcal P}'-\mathcal P'\mu_{\mathcal P} +2\mathcal C\mathcal \mu_{\mathcal C}'+\mathcal C'\mu_{\mathcal C}-2\mathcal H \mathcal \mu_{\mathcal H}'-\mathcal H'\mu_{\mathcal H}+\mu_{\mathcal H}'''\,,\\
\dot{ \mathcal{P}}&=2\mathcal P \mathcal \mu_{\mathcal L}'+\mathcal P'\mu_{\mathcal L}+2\mathcal H\mathcal \mu_{\mathcal C}'+\mathcal H'\mu_{\mathcal C}-\mu_{\mathcal C}'''-\frac{1}{\ell^2}\left(2\mathcal C \mathcal \mu_{\mathcal H}'+\mathcal C'\mu_{\mathcal H}+2\mathcal L \mathcal \mu_{\mathcal P}'+\mathcal L'\mu_{\mathcal P}\right)\,,\\
\dot{ \mathcal{C}}&=-2\mathcal H \mathcal \mu_{\mathcal P}'-\mathcal H'\mu_{\mathcal P}+\mu_{\mathcal P}'''+2\mathcal C \mathcal \mu_{\mathcal L}'+\mathcal C'\mu_{\mathcal L}\,,\\
\dot{ \mathcal{H}}&=2\mathcal H \mathcal \mu_{\mathcal L}'+\mathcal H'\mu_{\mathcal L}-\mu_{\mathcal L}'''-\frac{1}{\ell^2}\left(2\mathcal C \mathcal \mu_{\mathcal P}'+\mathcal C'\mu_{\mathcal P}\right)\,.
\end{align}

Transformation laws of the charge densities read
\begin{align}
\delta \mathcal{L}&=2\mathcal L \mathcal \varepsilon_{\mathcal L}'+\mathcal L'\varepsilon_{\mathcal L}-2\mathcal P\mathcal \varepsilon_{\mathcal P}'-\mathcal P'\varepsilon_{\mathcal P} +2\mathcal C\mathcal \varepsilon_{\mathcal C}'+\mathcal C'\varepsilon_{\mathcal C}-2\mathcal H \mathcal \varepsilon_{\mathcal H}'-\mathcal H'\varepsilon_{\mathcal H}+\varepsilon_{\mathcal H}'''\,,\\
\delta \mathcal{P}&=2\mathcal P \mathcal \varepsilon_{\mathcal L}'+\mathcal P'\varepsilon_{\mathcal L}+2\mathcal H\mathcal \varepsilon_{\mathcal C}'+\mathcal H'\varepsilon_{\mathcal C}-\varepsilon_{\mathcal C}'''-\frac{1}{\ell^2}\left(2\mathcal C \mathcal \varepsilon_{\mathcal H}'+\mathcal C'\varepsilon_{\mathcal H}+2\mathcal L \mathcal \varepsilon_{\mathcal P}'+\mathcal L'\varepsilon_{\mathcal P}\right)\,,\\
\delta \mathcal{C}&=-2\mathcal H \mathcal \varepsilon_{\mathcal P}'-\mathcal H'\varepsilon_{\mathcal P}+\varepsilon_{\mathcal P}'''+2\mathcal C \mathcal \varepsilon_{\mathcal L}'+\mathcal C'\varepsilon_{\mathcal L}\,,\\
\delta \mathcal{H}&=2\mathcal H \mathcal \varepsilon_{\mathcal L}'+\mathcal H'\varepsilon_{\mathcal L}-\varepsilon_{\mathcal L}'''-\frac{1}{\ell^2}\left(2\mathcal C \mathcal \varepsilon_{\mathcal P}'+\mathcal C'\varepsilon_{\mathcal P}\right)\,.
\end{align}
The algebra of the canonical generators can be directly obtained after using the identity
\begin{equation}
\lbrace Q_{\epsilon_{1}},Q_{\epsilon_{2}} \rbrace =\delta_{\epsilon_{2}}Q_{\epsilon_{1}}\,,
\end{equation}
and the transformation laws of the fields previously written. The Poisson bracket $\big\{\,,\,\big\}$, taken at the equal time slice $t=t_0$, between the charges are then given by
\begin{align}
i\left\{ \mathcal{L}_{m},\mathcal{L}_{n}\right\} 	&=	\left(m-n\right)\mathcal{L}_{m+n}\, , \\
i\left\{ \mathcal{L}_{m},\mathcal{P}_{n}\right\} 	&=	\left(m-n\right)\mathcal{P}_{m+n}\, , \\
i\left\{ \mathcal{P}_{m},\mathcal{P}_{n}\right\} 	&=	\frac{1}{\ell^{2}}  \left(m-n\right)\mathcal{L}_{m+n} \, , \\
i\left\{ \mathcal{L}_{m},\mathcal{H}_{n}\right\} 	&= \left(m-n\right)\mathcal{H}_{m+n}+km^{3}\delta_{m+n,0} \,, \\
i\left\{ \mathcal{L}_{m},\mathcal{C}_{n}\right\} 	&=	\left(m-n\right)\mathcal{C}_{m+n}\,, \\
i\left\{ \mathcal{P}_{m},\mathcal{H}_{n}\right\} 	&=	\frac{1}{\ell^{2}}  \left(m-n\right)\mathcal{C}_{m+n} \, , \\
i\left\{ \mathcal{C}_{m},\mathcal{P}_{n}\right\} 	&= 	\left(m-n\right)\mathcal{H}_{m+n}+km^{3}\delta_{m+n,0}\,,
\end{align}
with vanishing components
\begin{equation}
i\left\{ \mathcal{H}_{m},\mathcal{H}_{n}\right\}=i\left\{ \mathcal{C}_{m},\mathcal{C}_{n}\right\}=i\left\{ \mathcal{C}_{m},\mathcal{H}_{n}\right\}= 0 \,. \\
\end{equation}
Here, we have expanded in Fourier modes according to
\begin{equation}
X_{n}=\frac{k}{2\pi}\int d\phi X(\phi)e^{-in\phi} \,.
\end{equation}

This infinite-dimensional extension of the AdS$_3$ Carroll algebra contains four classes of generators, namely, spatial superrotations $\mathcal{L}_{m}$, Carroll superboosts $\mathcal{C}_m$, spatial supertranslations $\mathcal{P}_m$ and time supertranslations $\mathcal{H}_m$, possessing as subalgebra to the (centerless) conformal algebra in two dimensions, generated by the set $(\mathcal{L}_m,\mathcal{P}_n)$. The (centrally extended) BMS$_3$ algebra \cite{Barnich:2006av} is also included as a subalgebra spanned by $(\mathcal{L}_m,\mathcal{H}_n)$. We observe that all canonical generators transform in a $sl(2, R)$-representation of spin one (or conformal weight two). Notably, this algebra is isomorphic to the generalized BMS$_3$ algebra, previously obtained in \cite{Caroca:2017onr} through the semi-group expansion method by considering the action of the four-level cyclic semi-group $S^{(3)}_{M}$ (see e.g., \cite{Salgado:2013eut}) on BMS$_3$\footnote{We acknowledge P. Concha, E. Rodr\'iguez and D. Tempo  for this observation.}.

The wedge algebra (realized by restricting the generator labels as $m,n=\pm 1,0$) corresponds to the (doubled) extended AdS$_3$ Carroll algebra of \cite{Gomis:2019nih} (see also  \cite{Izaurieta:2009hz, Concha:2023bly}). In particular, the AdS$_3$ Carroll algebra is recovered with the following identification of the generators 
\begin{align}
J&=i\mathcal{L}_0\,,\qquad \qquad \quad \, H=\left(2\mathcal{H}_0+k\right)\,,\\
P_1&=\frac{i}{2}\left(\mathcal{P}_1+\mathcal{P}_{-1}\right)\,,\quad \, P_2=\frac{1}{2}\left(\mathcal{P}_1-\mathcal{P}_{-1}\right)\,, \\
C_1&=-i\left(\mathcal{C}_1-\mathcal{C}_{-1}\right)\,,\quad \,\,  C_2=-\left(\mathcal{C}_1+\mathcal{C}_{-1}\right)\,.
\end{align}

It is worth mentioning that boundary conditions, Lie-algebra gauge parameter, transformation laws of the dynamical fields, and charges for Carroll gravity in {the} absence of a cosmological constant, can be directly obtained by taking the flat limit ($\ell \rightarrow \infty$) in each step. Then, these results extend the previous ones in \cite{Bergshoeff:2016soe} by consistently incorporating spatial superrotations and time supertranslations in the asymptotic symmetry group. Interestingly, in the flat limit of the asymptotic symmetry algebra written above,  the subset formed by Carroll super boosts $\mathcal C_m$, spatial supertranslations $\mathcal P_n$ and time supertranslations $\mathcal H_n$ become an Abelian ideal. In consequence, the two-dimensional conformal subalgebra (generated by spatial superrotations $\mathcal L_m$ and spatial supertranslations $\mathcal P_n$) reduces in the flat limit to the (centerless) BMS$_3$ algebra, where (as expected) all supertranslations commute each other. The asymptotic symmetry algebra in this case is isomorphic to the generalized BMS$_3$ algebra obtained from the action of the five-level semi-group $S^{(3)}_{E}$ (see e.g., \cite{Izaurieta:2006zz}) on BMS$_3$.

\subsection{Heisenberg current algebra, BMS$_3$ and composite $W(0,-s)$ generators} \label{sec:Heisenberg}
It is also possible to introduce boundary conditions that turn out to be preserved by an Abelian group whose canonical generators satisfy the Heisenberg current algebra. Configurations contained within this set of boundary conditions could be considered as Carrollian analogues of the so-called soft hairy black holes and cosmological configurations found in the context of three-dimensional Einstein gravity \cite{Afshar:2016wfy,Afshar:2016kjj}. 

We make use of the auxiliary components of the gauge connection in \cite{Afshar:2016wfy,Afshar:2016kjj}, which, in terms of AdS Carroll generators, are given by
\begin{subequations}\label{eq:Heisenbergconn}
\begin{align}
a_{\phi}&=\mathcal{E}J+\mathcal{J}H\,,\\
a_t&=\mu_{\mathcal J}J+\mu_{\mathcal E}H\,,
\end{align}
\end{subequations}
where $\mathcal E$ and $\mathcal J$ are dynamical fields (arbitrary functions of $t$ and $\phi$), while the chemical potentials $\mu_{\mathcal J}$ and $\mu_{\mathcal E}$ (arbitrary functions of $t$ and $\phi$ as well) are left fixed without variation, i.e., $\delta \mu_{\mathcal J}=\delta \mu_{\mathcal E}=0$. {Here the generators $J$ and $H$ satisfy the AdS Carroll algebra in \eqref{eq:AdSCarrollalgebra}}. {The gauge connection \eqref{eq:Heisenbergconn}} is preserved by the action of the gauge parameter
\begin{equation}
\lambda=\varepsilon_{\mathcal J}J+\varepsilon_{\mathcal E}H\,,
\end{equation}
provided transformation laws of the fields read
\begin{equation}
\delta \mathcal E=\varepsilon_{\mathcal J}'\,,\quad \delta \mathcal J=\varepsilon_{\mathcal E}'\,.
\end{equation}
The corresponding global charge is given by the following surface integral
\begin{equation}
\delta Q=-\frac{k}{2\pi}\oint d\phi \, \left(\varepsilon_{\mathcal J}\delta \mathcal J+\varepsilon_{\mathcal E}\delta \mathcal E\right)\,.
\end{equation}
The algebra of the canonical generators of the asymptotic symmetries reads
\begin{align}
i\left\{ \mathcal J_{m},\mathcal J_{n}\right\} 	&=	0\,,\\
i\left\{ \mathcal E_{m},\mathcal E_{n}\right\} 	&=	0\,,\\
i\left\{ \mathcal J_{m},\mathcal E_{n}\right\} 	&=	km \delta_{m+n,0}\,.
\end{align}
This algebra is isomorphic to two affine $\hat{u}(1)$ current algebras with the same levels $k/2$. As shown in \cite{Afshar:2016wfy,Afshar:2016kjj}, there exists a change of basis for the canonical generators that permits to write  these commutators in Casimir-Darboux coordinates, which leads to the Heisenberg current algebra (see, e.g. Section VI of \cite{Afshar:2016wfy}). From here, it is then direct to show that one can recover the (centrally extended) BMS$_3$ algebra,
\begin{align}
i\left\{ \mathcal L_{m},\mathcal L_{n}\right\} 	&=	(m-n)\mathcal L_{m+n}\,,\\
i\left\{ \mathcal L_{m},\mathcal P_{n}\right\} 	&=	(m-n)\mathcal P_{m+n}+km^3\delta_{m+n,0}\,,\\
i\left\{ \mathcal P_{m},\mathcal P_{n}\right\} 	&=	0\,,
\end{align}
from the twisted Sugawara construction
\begin{align}
\mathcal L&= \mathcal E \mathcal J+\mathcal J'\,,\\
\mathcal P&= \frac{1}{2}\mathcal E^2+\mathcal E'\,.
\end{align}

{Going further,} one can also construct composite $W(0,-s)$ generators, which satisfy the algebra
\begin{align}
i\left\{ \mathcal L_{m},\mathcal L_{n}\right\} 	&=	(m-n)\mathcal L_{m+n}\,,\\
i\left\{ \mathcal L_{m},\mathcal P^{(s)}_{n}\right\} 	&=	(s\,m-n)\mathcal P^{(s)}_{m+n}\,,\\
i\left\{ \mathcal P^{(s)}_{m},\mathcal P^{(s)}_{n}\right\} 	&=	0\,,
\end{align}
by considering the following Sugawara construction,
\begin{equation}
\mathcal L= \mathcal E \mathcal J\,,\qquad
\mathcal P^{(s)}= \mathcal E^{(s+1)}\,.
\end{equation}
The $W(0,b)$ algebra belongs to a more general class of algebras denoted by $W(a,b)$. This corresponds to an extension of the de Witt algebra \cite{Ovsienko:1994im,Roger:2006rz,Gao:2011}, which can also be obtained from algebraic deformations of BMS$_3$ \cite{FarahmandParsa:2018ojt}. In the case $a=0$ and $b=-s$, the generator $\mathcal P^{(s)}_n$ has arbitrary conformal weight $h=s+1$, it can then be regarded as a supertranslation generator of spin $s$. This sort of extension has been found as near-horizon asymptotic symmetry algebras in  \cite{Donnay:2015abr,Donnay:2016ejv} for $s=0$ and in \cite{Grumiller:2019fmp} for arbitrary $s$, while the case $s=1$ corresponds to the (centerless) BMS$_3$ algebra. Further aspects related to the BRST quantization of field theories associated {with} this one-parameter class of algebras (also known as BMS-like field theories) have been carried out in \cite{Figueroa-OFarrill:2024wgs}.

\section{Carroll thermal configuration}
This section is devoted to the study of the Carroll structures, isometries, global charges and Carroll thermal properties of a new solution fulfilling the boundary conditions introduced in Section \ref{sec:asymptotics}.
\subsection{Solution: Carroll structures, global charges and Carroll extremal surface} \label{subsec:Configuration}
For the sake of simplicity, we will consider configurations endowed with constant phase space fields $\mathcal J$, $\mathcal M$, $\mathcal N$ and $\mathcal S$. Under this consideration equations of motion \eqref{eq:eqmotion1}-\eqref{eq:eqmotion2} hold provided  $\mu_{\mathcal P}=0$ and the remaining chemical potentials $\mathcal{\mu}_{\mathcal J}$, $\mathcal{\mu}_{\mathcal H}$ and $\mathcal{\mu}_{\mathcal C}$ take constant values.

Carroll structures describing the solutions within the set of boundary conditions in Section \ref{sec:asymptotics} are obtained by consistently incorporating the radial coordinate. The latter will be realized through a permissible gauge transformation 
\begin{equation}
A(t,r,\phi)=b^{-1}a(t,\phi)b+b^{-1}\,db\,,
\end{equation} 
on the auxiliary connection $a$, with nonvanishing components in \eqref{eq:aphi} and \eqref{eq:at}, generated by the group element
\begin{equation}
b(r)=e^{\alpha(r) P_{2}}\qquad \text{with}\qquad \alpha(r)=\ell \sinh^{-1}\left(\frac{r}{\ell}\right) \,.
\end{equation}
The gauge connection then reads
\begin{align}\label{eq:A}
A=&\left(e^{\mathcal{S}}f(r)-\frac{\mathcal N r}{\ell^2}\right)(d\phi+\mu_{\mathcal{L}}dt)J-\left(e^{\mathcal S}r-\mathcal Nf(r)\right)(d\phi+\mu_{\mathcal{L}}dt)P_1+\frac{dr}{f(r)}P_2\nonumber\\
&+ \mathcal M(d\phi+\mu_{\mathcal{L}}dt)C_1+\left[\frac{\mathcal J r}{\ell^2}(d\phi+\mu_{\mathcal{L}}dt)+\frac{\mu_{\mathcal H}}{\ell^2}\left(e^{\mathcal S}r-\mathcal Nf(r)\right)dt\right]C_2 \nonumber\\
&+\left[\mathcal J f(r)(d\phi+\mu_{\mathcal{L}}dt)+\mu_{\mathcal H}\left(e^{\mathcal S}f(r)-\frac{\mathcal N r}{\ell^2}\right)dt\right]H\,,
\end{align}
where
\begin{equation}
f^2(r)=\frac{r^{2}}{\ell^2}+1\,.
\end{equation}
We can easily read off -- recalling \eqref{eq:gauge-connection} -- the vielbeins $e^a$ and $\tau$ , which go along the spatial and time translation generators, respectively. {Performing the change of frame $\phi\rightarrow \phi-\mu_{\mathcal L}t$, the singular line element is then given by}
\begin{equation}
ds^2=\delta_{ab}\, e^a_{\mu} \, e^b_{\nu}\, dx^{\mu} dx^{\nu}=\frac{dr^2}{f^2(r)}+\left(e^{\mathcal S}r-\mathcal Nf(r)\right)^2d\phi^2\,,
\end{equation}
while the temporal vielbein reads  
\begin{equation}
\tau=\mu_{\mathcal H}\left(e^{\mathcal S}f(r)-\frac{\mathcal N r}{\ell^2}\right)dt+\mathcal J f(r)d\phi\,.
\end{equation}
As explained in Section \ref{sec:action}, the remaining Carroll structures can be unambiguously determined through the completeness conditions \eqref{eq:completeness}. Hence, the normal vector and the components of the inverse spatial vielbeins are given by
\begin{equation}\label{eq:normal-vector}
n^{\mu}=\left(\frac{1}{\displaystyle \mu_{\mathcal H}\left(e^{\mathcal S}f(r)-\frac{\mathcal N r}{\ell^2}\right)},0,0\right)\,,
\end{equation}
and
\begin{align}
E^{\mu}_1&=\left(\frac{\mathcal J f(r)}{\left(e^{\mathcal S}f(r)-\frac{\mathcal N r}{\ell^2}\right)\left(e^{\mathcal S}r-\mathcal Nf(r)\right)},0,-\frac{1}{\left(e^{\mathcal S}r-\mathcal Nf(r)\right)}\right)\,,\\
E^{\mu}_2&=\left(0,f(r),0\right)\,,
\end{align}
respectively.

Some comments on this solution are in order:

\begin{itemize}
\item The diffeomorphisms that preserve the Carroll structure for this solution, namely, 
\begin{equation}
\mathcal L_{\xi}g_{\mu\nu}=\mathcal L_{\xi}n^{\mu}=0\qquad \text{with}\qquad \xi=\xi^{\mu}\partial_{\mu}\,,
\end{equation}
are generated by the infinite-dimensional set of Killing vectors $\Xi(r,\phi)\partial_t$ and $\partial_{\phi}$, where $\Xi(r,\phi)$ is an arbitrary function of the radius and the angular coordinate. The emergence of infinite-dimensional symmetries is indeed not surprising, but it seems to be a sign of the degeneracy proper of a Carroll geometry. A similar fact stands for the case of a flat Carroll geometry in $D$ dimensions, which is left invariant by the infinite-dimensional Carroll group $\mathcal{C}(D)$ \cite{Henneaux:1979vn} (see also \cite{Henneaux:2021yzg}). 

\item The global charges associated to this class of configurations are obtained from \eqref{eq:charge}, where $\lambda=-\xi^{\mu}A_\mu$ (setting $\mu_{\mathcal H}=-1$ and $\mu_{\mathcal L}=0$), they correspond to the energy $E=Q[\partial_t]$ and the angular momentum $L=Q[\partial_{\phi}]$. From the gauge connection \eqref{eq:A}, we find that
\begin{align}
E&=k\mathcal H=\frac{k}{2}\left(\frac{\mathcal{N}^{2}}{\ell^{2}}-e^{2\mathcal{S}}\right)\,, \label{eq:E}\\
L&=k \mathcal L=k e^{\mathcal{S}}\mathcal{J}\,,\label{eq:L}
\end{align}
where we have made use of the explicit expressions for the charge densities $\mathcal L$ and $\mathcal H$ in \eqref{eq:mathcalL} and \eqref{eq:mathcalH}, respectively. These charges are measured with respect to the energy of the AdS$_3$ Carroll geometry $E=-k/2$. In these coordinates, the AdS$_3$ Carroll geometry is described by the following Carroll structures
\begin{equation}\label{eq:AdS-CarrollGeo}
ds^2=\frac{dr^2}{\frac{r^2}{\ell^2}+1}+r^2d\phi^2\,,\quad n^{\mu}=\left(-\frac{1}{\sqrt{\frac{r^2}{\ell^2}+1}},0,0\right)\,,
\end{equation}
which can be directly obtained by turning off all integration constants. Note that similarly to the flat Carroll geometry,  \eqref{eq:AdS-CarrollGeo} is left invariant by an infinitesimal-dimensional set of diffeomorphisms generated by a vector field $\xi^{\mu}$, with components
\begin{align}
\xi^t&=\Xi(r,\phi)-\frac{rt}{\ell f(r)}\left(W_1 \cos(\phi)+W_2 \sin(\phi)\right)\,,\\
\xi^r&=\ell f(r)\left(W_1 \cos(\phi)+W_2 \sin(\phi)\right)\,,\\
\xi^{\phi}&=\frac{\ell f(r)}{r}\left(W_1 \sin(\phi)-W_2 \cos(\phi)\right)+Y\,,
\end{align}
where $\Xi(r,\phi)$ is again arbitrary. The constants $W_1$, $W_2$ and $Y$ parametrize the kinematical subgroup of transformations $SO(2,1)$, while the function $\Xi(r,\phi)$ is related to the dynamical ones. In fact, in the case of choosing the function $\Xi(r,\phi)$ as
\begin{equation}
\Xi(r,\phi)=T-\frac{r}{f(r)}\left(B_1 \cos(\phi)+B_2 \sin(\phi)\right)\,,
\end{equation} 
the Lie bracket for this vector field forms a representation the AdS$_3$ Carroll algebra, where $T$ stands for the time translation parameter, while $B_1$ and $B_2$ generate Carroll boosts. For this reason, the vector field $\xi^{\mu}$ (with an arbitrary function $\Xi(r,\phi)$) can be regarded as the generator of AdS-$\mathcal C (3)$, the AdS extension of the infinite-dimensional Carroll group $\mathcal C (3)$.\\

\item The solution seems to possess a degeneracy in its description (it is characterized by only two global charges, energy and angular momentum). Infinite degeneracies in the spectrum of Carroll field theories are argued to arise due to their ultralocality in the quantization process  \cite{deBoer:2023fnj}. This case would be an example of a finite degeneracy in a classical realization of Carroll gravity. However, we must note that the parameter $\mathcal{M}$ can be set to zero (in the solution) by a permissible gauge transformation generated by an appropriate group element along $C_2$ (see subsection \ref{sec:thermo} below). It would be desirable to discard the possibility of eliminating the whole degeneracy by employing suitable permissible gauge transformations. A throughout analysis on this aspect will be addressed elsewhere.

\item The normal vector $n^{\mu}$ in \eqref{eq:normal-vector} diverges at {the radius}
\begin{equation}\label{eq:r0}
r_{0}=\frac{e^{\mathcal S}}{\sqrt{\frac{\mathcal{N}^2}{\ell^2}-e^{2\mathcal S}}}\,,
\end{equation}
provided $\mathcal{N}>0$ and $\frac{\mathcal{N}^2}{\ell^2}-e^{2\mathcal S}>0$, which amounts to ask for $E>0$. It is then concluded that under the latter conditions, the configuration is endowed with a Carroll extremal surface at $r=r_0$, defined in \cite{Ecker:2023uwm}. We can now make use of the Carrollian analogue of the Euclidean continuation of the solution in order to deepen into the C-thermal properties of this manifold. 

\item Note that in the flat limit $(\ell \rightarrow \infty)$, the normal vector \eqref{eq:normal-vector} becomes constant, then the resulting flat configuration is devoid of a Carrollian structure singularity. Nonetheless, it would be certainly interesting to explore different sets of boundary conditions (if any) that allow to define three-dimensional flat C-thermal configurations.
\end{itemize}

\subsection{C-thermal properties: thermal holonomy and entropy}\label{sec:thermo}
The gauge field formulation of the theory allows us to  address the characterization of the configurations contained in these boundary conditions. Thus, by taking advantage of the tools developed in the context of higher spin extensions of three-dimensional gravity, along the lines of  \cite{Henneaux:2013dra,Bunster:2014mua}, we will ask for regularity of the Carroll analogue of the Euclidean continuation of the solution by demanding the topology of a solid torus. The solution is manifestly endowed with arbitrary chemical potentials, then for the analysis below, the fixed ranges of the Euclidean time and the angular coordinate reads 
\begin{equation}
0\leq \tau < 1 \quad \text{and} \quad 0\leq \phi < 2\pi\,,
\end{equation}
respectively.

In what follows, we will determine the configuration with trivial holonomy along a thermal cycle. This procedure will fix the chemical potentials in terms of the global charges. We will then proceed to compute the entropy associated with this configuration by employing the Chern-Simons entropy formula in \cite{Perez:2012cf,Perez:2013xi}, further developed in \cite{deBoer:2013gz,Bunster:2014mua}
\begin{equation}\label{eq:CSformula}
S=\frac{k}{2\pi}\oint d\phi \langle A_{t}A_{\phi} \rangle_{\text{on-shell}}\,.
\end{equation} 
It is then explicitly verified that this solution does satisfy the first law of thermodynamics, i.e., it is a C-thermal manifold in the sense of \cite{Ecker:2023uwm}. Nonetheless, it will be argued below that the configuration actually corresponds to a Carroll analogue to the cosmological configurations that arise in the context of asymptotically flat spacetimes for three-dimensional (higher spin) gravity theories \cite{Ezawa:1992nk,Cornalba:2002fi,Cornalba:2003kd,Barnich:2012xq,Gary:2014ppa,Matulich:2014hea}.

Smoothness of the regular solutions is implemented by requiring that the holonomy of the gauge field must be contractible along a thermal cycle $C$. This amounts to {asking} for the condition
\begin{equation}\label{eq:holoeq}
P e^{\int_C a_{\mu}dx^{\mu}}=e^{\int_0^1 a_{\tau}d\tau}=\Gamma\,,
\end{equation}
where $\Gamma$ belongs the center of the AdS$_3$ Carroll group. We note at this point that we do not account with a suitable matrix representation from which the invariant bilinear product in \eqref{eq:bracket} can be obtained from the trace of the product of two generators.\footnote{The diagonalization of the holonomy could be implemented through a nonstandard matrix representation in \cite{Krishnan:2013wta} by making use of the isomorphism of AdS Carroll and the (para-)Poincar\'e algebra.} Then, we implement in this case the procedure introduced in \cite{Matulich:2014hea}, which is realized by applying a permissible group element of the form $g=e^{\lambda_1 C_1+\lambda_2 C_2}$. This will permit to gauge away the components along $C_a$ and $H$, and then using a suitable matrix representation (for the kinematical generators $J$ and $P_a$) to diagonalize the holonomy.

From equation \eqref{eq:at},  it is direct to check that, after applying the gauge transformation with the group element $g=e^{\lambda_1 C_1+\lambda_2 C_2}$ (being $\lambda_1$ and $\lambda_2$ phase space functions to be determined), the time component of the auxiliary connection associated with the solution
\begin{equation}
a_t=e^{\mathcal{S}}\mu_{\mathcal{L}}J-\frac{1}{\ell^2}\mathcal N \mu_{\mathcal H} C_2+\left(\mathcal J \mu_{\mathcal L}+e^{\mathcal S}\mu_{\mathcal H} \right)H+\mathcal M \mu_{\mathcal L}C_1+\mathcal N \mu_{\mathcal L}P_1\,,
\end{equation}
becomes
\begin{align}\label{eq:attransformed}
a_t&=e^{\mathcal{S}}\mu_{\mathcal{L}}J+\mathcal N \mu_{\mathcal L}P_1+\left(\lambda_1 e^{\mathcal S}\mu_{\mathcal L}-\frac{1}{\ell^2}\mathcal N \mu_{\mathcal H}\right) C_2\nonumber\\
&\quad+\left(\mathcal J \mu_{\mathcal L}+e^{\mathcal S}\mu_{\mathcal H} -\lambda_1 \mathcal N \mu_{\mathcal L}\right)H+\left(\mathcal M-e^{\mathcal S}\lambda_2 \right) \mu_{\mathcal L}C_1\,.
\end{align}
As mentioned, we aim to eliminate the terms along the generators associated to dynamical transformations, namely, $H$ and $C_a$. Components along Carroll boost generators vanish provided
\begin{equation}
\lambda_1=\frac{\mathcal N\mu_{\mathcal H}e^{-\mathcal S}}{\ell^2\mu_{\mathcal L}}\quad \text{and} \quad \lambda_2=e^{-\mathcal S}\mathcal M \,.
\end{equation}
The vanishing of the term along the time translation generator $H$ leads to the relation
\begin{equation}\label{eq:muH}
\mu_{\mathcal H}=\frac{ e^{\mathcal S}\mathcal J \mu_{\mathcal L}}{\frac{\mathcal N^2}{\ell^2}-e^{2\mathcal S}}=\frac{\mu_{\mathcal L}\mathcal L}{2 \mathcal H}\,,
\end{equation}
where we have used the expression for charge densities in \eqref{eq:E} and \eqref{eq:L}. The time component of the gauge connection $a_t$ {in \eqref{eq:attransformed}} then reduces to
\begin{equation}
a_t=e^{\mathcal{S}}\mu_{\mathcal{L}}J+\mathcal N \mu_{\mathcal L}P_1\,,
\end{equation}
which only possesses terms along the generators of the kinematical transformations. By direct use of the {faithful} $2\times 2$ matrix representation
\begin{equation}
J=\left(\begin{array}{cc}
0 & -\frac{1}{2}\\
\frac{1}{2} & 0
\end{array}\right)\,,\quad P_1=\left(\begin{array}{cc}
0 & -\frac{1}{2\ell}\\
-\frac{1}{2\ell} & 0
\end{array}\right)\,,\quad P_2=\left(\begin{array}{cc}
\frac{1}{2\ell} & 0\\
0 & -\frac{1}{2\ell}
\end{array}\right)\,,
\end{equation}
we find that the holonomy around a thermal cycle is trivial provided
\begin{equation}\label{eq:muL}
\mu_{\mathcal L}=\frac{2n\pi}{\sqrt{\frac{\mathcal N^2}{\ell^2}-e^{2\mathcal S}}}\,,
\end{equation}
where the  center group element is given by $\Gamma=-\mathbb{I}_2$ if $n$ is an odd integer, while $\Gamma=\mathbb{I}_2$ as long as $n$ is an even integer. In this case we will choose $n=\pm 1$, where the sign is related to spin direction of the solution $\text{sgn}(\mathcal L)$. The chemical potentials are then given by
\begin{equation}
\mu_{\mathcal L}=\text{sgn}(\mathcal L)\pi\sqrt{\frac{2}{\mathcal H}}\,,\qquad \mu_{\mathcal H}=\frac{\pi|\mathcal L|}{\sqrt{2}\mathcal H^{\frac{3}{2}}}\,.
\end{equation}

By direct application of the Chern-Simons formula \eqref{eq:CSformula}, we find that the entropy of the configuration in terms of the global charges reads
\begin{equation}\label{eq:S}
S=2k\left( \mu_{\mathcal L}\mathcal L-\mu_{\mathcal H}\mathcal H\right)=\pi | L|\sqrt{\frac{2k}{E}}\,.
\end{equation}
Note that the entropy is positive and real provided the energy $E$ in \eqref{eq:E} is strictly positive (or equivalently $\frac{\mathcal N^2}{\ell^2}-e^{2\mathcal S}>0$), which amounts to ensure the existence of the Carroll extremal surface in \eqref{eq:r0}. Thus, sensible C-thermal properties go hand in hand with the existence of the Carroll extremal surfaces introduced in \cite{Ecker:2023uwm}.

From the entropy in \eqref{eq:S}, we can now readily obtain the temperature $T$ and chemical potential for the angular momentum $\Omega$ (angular velocity)

\begin{align}
\beta&=\left(\frac{\partial S}{\partial E}\right)_{L}=-\pi |L|\sqrt{\frac{k}{2E^3}}\,,\\
\beta \, \Omega&= -\left(\frac{\partial S}{\partial L}\right)_{E}=-\text{sgn}(L)\pi\sqrt{\frac{2k}{E}}\,.
\end{align}
These expressions reveal that $\beta$ is indeed negative, which is a sign that we might be dealing with a Carrollian analogue of a cosmological configuration (reversed orientation of the solid torus as compared with a black hole configuration. For more details see \cite{Matulich:2014hea}). By following this proposal, we consider that $\beta=-1/T$, then the temperature and angular velocity are given by
\begin{equation}
T=\frac{1}{\pi |L|}\sqrt{\frac{2E^3}{k}}\quad\text{and}\quad\Omega=\frac{2E}{L}\,,
\end{equation}
respectively. It is then straightforward to verify that the first law of thermodynamics $\delta S=\beta \delta E-\beta \Omega \delta L$ holds in the grand canonical ensemble (with $E$ and $L$ given in \eqref{eq:E} and \eqref{eq:L}, respectively) for this C-thermal manifold.

It is worth highlighting that the form of the entropy and the chemical potentials for this C-thermal configuration (in terms of its global charges) resembles the ones of the asymptotically flat cosmological configuration of three-dimensional Einstein gravity  \cite{Barnich:2012xq,Gary:2014ppa,Matulich:2014hea}. This intriguing fact indicates a non-trivial connection of the solution in this Carrollian set-up with the one of the Lorentzian case, although the features of the solutions strongly differs (marked by the degeneracy of the Carrollian configuration). The isomorphism between AdS-Carroll and Poincar\'e symmetries might be crucial to further elaborate and understand this point (see e.g. \cite{Aviles:2024llx}). We will not search for an explanation for this curiosity here, but it will be left for future explorations.

\section{Concluding remarks}\label{sec:endings}

In this article, we have proposed a new set of boundary conditions in the context of AdS$_3$ Carroll gravity. The asymptotic symmetries of the theory were shown to be described by an infinite-dimensional extension of the AdS$_3$ Carroll group, parametrized by four functions of the circle at infinity, which is isomorphic to a precise generalized BMS$_3$ algebra (previously obtained in \cite{Caroca:2017onr} by algebraic expansion methods). We have also introduced a novel three-dimensional C-thermal manifold (in the sense of \cite{Ecker:2023uwm}) and explored its thermal properties by taking advantage of the gauge field formulation of the theory and the interesting concepts introduced in \cite{Ecker:2023uwm}. By assuming the topology of a solid torus, regularity requirements (reflected in a trivial holonomy condition along the thermal cycle) fixed the value of the chemical potentials in terms of the global charges and proceed to compute the entropy of the configuration by making use of the Chern-Simons formula developed in the context of three-dimensional higher spin black holes \cite{Perez:2012cf,Perez:2013xi} (see also e.g., \cite{deBoer:2013gz,Bunster:2014mua}).

We have also shown that it is possible to write boundary conditions in the context of AdS$_3$ Carroll gravity, whose asymptotic symmetry algebra is given by the Heisenberg current algebra. It would be then interesting to delve into possible Carroll analogues of soft hairy black holes \cite{Afshar:2016wfy}, and realize the analysis of their C-thermal properties.

The solution and boundary conditions introduced in this work were formulated from an intrinsic Carrollian viewpoint. It is certainly interesting to explore whether these results can be obtained from a gravitational ancestor, i.e., by taking a suitable process limit or exploiting the isomorphism between the AdS-Carroll and Poincaré algebras, and compare with the existing results extending the Brown-Henneaux boundary conditions \cite{Brown:1986nw} in three-dimensional Einstein gravity \cite{Compere:2013bya,Troessaert:2013fma,Avery:2013dja,Afshar:2016wfy,Grumiller:2016pqb,Afshar:2016kjj,Detournay:2016sfv,Grumiller:2017sjh}.

A rigorous analysis of the solution space within this new set of boundary conditions is worth performing as well. This can be done through an exhaustive study of the Wilson loops for both Carroll spin connections and the AdS-Carroll gauge field, and exploring the question on the existence of central singularities in the case of Carroll geometries along the lines of \cite{Briceno:2024ddc}.

\section*{Acknowledgements}
We thank {Patrick Concha, Joaquim Gomis, Matthias Harksen, Marc Henneaux,  Sucheta Majumdar, Niels Obers, Evelyn Rodr\'iguez, David Tempo, Ricardo Troncoso and Jorge Zanelli} for useful comments and many interesting discussions. D.H. and P.R. are grateful to the ``Physique Th\'eorique et Math\'ematique''  group at ULB for kind hospitality, where this collaboration started. O.F. is grateful to Prof. L\'arus Thorlacius and Science Institute of the University of Iceland for warm hospitality, where part of this work was carried out. O.F. thanks the organizers of the ESI Programme and Workshop ``Carrollian Physics and Holography'' hosted by the Erwin Schr\"odinger Institute in April 2024 in Vienna, where part of this work was completed. D.H. is supported by the Icelandic Research Fund Grant 228952-053 and by the University of Iceland Research Fund. This research has been partially supported by ANID through Fondecyt grants N$^\circ$ 11251195, 3220805, 3230633 and SIA grant N$^\circ$ 85220027. This work was partially supported by a Marina Solvay Fellowship (O.F.) and by FNRS-Belgium (IISN 4.4503.15), as well as by funds from the Solvay Family. The research of O.F. is also partially supported by the Vicerrector\'{i}a de Investigaci\'on e Innovaci\'on of the Universidad Arturo Prat through an UNAP Consolida grant.

%--------------------------------------------------------
%------------------------------------

\end{document}